\documentstyle[epsf,11pt]{article}
\begin{document}

\begin{titlepage}

\begin{flushright}
   TPR-98-16
\end{flushright}

\vspace{1.5cm}

\begin{center}
{\Large\bf Fine-Tuning Two-Particle Interferometry\newline
II: Opacity Effects}
\end{center}

\vspace{1cm}

\begin{center}
 {\large Boris Tom\'a\v{s}ik and Ulrich Heinz}
\end{center}

\vspace{0.5cm}

\begin{center}
   {\it    
   Institut f\"ur Theoretische Physik, Universit\"at Regensburg,\\
   D-93040 Regensburg, Germany}

\vspace{2.5cm}

{May 7, 1998}
\end{center}

\vspace{1.1cm}

\abstract
We present a model study of single-particle spectra and two-particle
Bose-Einstein correlations for opaque sources. We study the transverse
mass dependence of the correlation radii $R_\perp$, $R_\parallel$ and
$R_0$ in the YKP parametrization and find a strong sensitivity of the
temporal radius parameter $R_0^2$ to the source opacity. A simple
comparison with the published data from 158~$A$~GeV/$c$ Pb+Pb
collisions at CERN indicates that the pion source created in these
collisions emits particles from the whole reaction volume and is 
not opaque. For opaque sources we find certain regions of 
inapplicability of the YKP parametrization which can be avoided by 
a slightly different parametrization for the correlator. The physical 
meaning of the modified parameters is briefly discussed. 

\endabstract
\vspace{2.5cm}

\end{titlepage}

\section{Introduction}
\label{intro}

Bose-Einstein correlations in the two-particle momentum spectra of
identical particle pairs provide a powerful tool to obtain information 
about the space-time structure of the particle emitting source. In a
previous paper \cite{th1} (referred to in the following as ``Paper
I'') we began a detailed numerical model study of two-particle 
Bose-Einstein correlations for ultrarelativistic heavy ion collisions.
We showed that collective expansion and temperature gradients lead
to an $M_\perp$ dependence of the correlation radii, and that by
studying this dependence one can reconstruct the dynamical state of
the collision fireball at freeze-out. 

A different feature of the particle source which can also cause
observable effects in the $M_\perp$ dependence of the correlation
radii but was not touched upon in Paper I is the source ``opacity''
\cite{HV96,HV97}. This mechanism, which could affect the geometric and
dynamical interpretation of the correlation radii, will be
investigated here. 

``Opaque'' sources emit particles from a thin shell near the fireball 
surface and are thus characterized by a smaller spatial extension of
the emission zone in the ``outward'' ($x$) than in the ``sideward''
($y$) direction\footnote{The tilde notation is defined in
  Sec.~\ref{model}; it indicates the variances of the distribution of
  emission points in space-time.}: $\langle \tilde x^2 - \tilde
y^2\rangle <0$. This is true in particular \cite{SH97} for emission
functions from hydrodynamical simulations where freeze-out is
implemented along a sharp hypersurface, characterized e.g. by a
constant freeze-out temperature. Transparent sources of the type
studied, for example, in \cite{CL96a,CNH95,WHTW96}, on the other hand,
feature a positive and generally small difference $\langle \tilde x^2
-\tilde y^2\rangle$ \cite{WHTW96}.

In Refs.~\cite{HV96,HV97} it was pointed out that this feature of 
opaque sources generally leads to smaller values for $R_o$ than for 
$R_s$ in the Cartesian parametrization of the correlator. In particular
for pairs with vanishing transverse momentum $K_\perp=0$, one has 
\cite{CSH95a} $R_s^2 = \langle \tilde y^2 \rangle > R_o^2 =
\langle \tilde x^2 \rangle$ if $\langle \tilde x^2 - \tilde y^2\rangle
<0$. The existence of a thin emission layer with directed emission
only into the outward hemisphere thus breaks the usual symmetry argument
\cite{CNH95,WSH96} that $\lim_{K_\perp\to 0} (R_o^2-R_s^2) = 0$. That 
argument is based on the assumed azimuthal symmetry of the effective 
source for vanishing $K_\perp$ where the direction of the transverse
pair momentum no longer serves to distinguish between the outward and
sideward directions (parallel and perpendicular to $\bf{K}_\perp$); for
opaque sources the orientation of the emitting surface itself provides
that distinction.

Here we show that in the Yano-Koonin-Podgoretski\u \i \ (YKP)
parametrization \cite{CNH95,HTWW96a} the opacity effects get enhanced
in the ``temporal'' radius parameter $R_0^2$ which turns negative for
small $K_\perp$ and diverges to minus infinity in the limit
$K_\perp\to 0$ if $\langle \tilde x^2 - \tilde y^2\rangle <0$. While
this would destroy the interpretation of $R_0$ in terms of an
effective source lifetime, it would provide particularly clear
evidence for surface dominated emission. Since no such evidence is
seen in the data from the NA49 Collaboration \cite{NA49corr,NA49QM97},
rather stringent limits on the degree of ``opaqueness'' of the source
created in these collisions can already now be established (see
Sec.~\ref{compar}).  

In the course of this study we discovered that for opaque sources
the YKP parame\-tri\-zation may become ill-defined in certain kinematic 
regions. In Sec.~\ref{mykp} we introduce a modification of the YKP 
parametrization without this defect. The cost for the remedy is a less 
straightforward physical interpretation of its radius parameters. 
Fortunately, for transparent sources the YKP parametrization with its
simpler space-time interpretation generally appears to work well.

For the general formalism and notation we refer the reader to 
Paper I. In the next Section we introduce the modification of the 
emission function from Paper I which is needed to parametrize the
opacity of the source. In Sec.~\ref{modelstudy} we describe the
results of our model study. Some problems with the YKP parametrization
for  opaque sources are discussed and corrected in
Sec.~\ref{modykp}, and the results are summarized in
Sec.~\ref{conclusions}. As in Paper I we consider only directly
emitted pions from the thermalized source, neglecting pions from
resonance decays whose effects were studied in Refs.~\cite{WH97}.

\section{A model for opaque sources}
\label{model}

Following the idea of Heiselberg and Vischer \cite{HV96,HV97} we
introduce the opacity into the model emission function from Paper I
via an additional exponential factor $\exp[ - \l_{\rm eff} / \lambda
]$ which suppresses the emission of the particles from deep inside the
source. $l_{\rm eff}(r,\phi)$ is the effective length which a particle
emitted at point $(r,\varphi)$ travels in outward ($x$) direction
before leaving the source. We implement the Gaussian transverse  
density profile as follows:
 \begin{equation}
 \label{mod1}
   l_{\rm eff}(r,\phi) = 
     e^{- \frac{y^2}{2R^2}} \, \int_x^{\infty} \,
     e^{- \frac{{x'}^2}{2R^2}} \, dx' \qquad {\rm with  } \quad
   y = r\, \sin \phi ,\: x = r\, \cos\phi\, .
 \end{equation} 
$\lambda$ represents the specific mean free path of the particle in
the medium. The actual mean free path would be obtained by dividing
$\lambda$ by the medium density; this is, however, already included in
Eq.~(\ref{mod1}). In the limit $\lambda \to \infty$ the transparent
source of Paper I is recovered. Note that, in contrast to \cite{HV97},
our mean free path $\lambda$ and transverse geometric radius $R$ are
time-independent. For later convenience we introduce the opacity
parameter 
 \begin{equation}
 \label{opacity}
    \omega = {R\over\lambda}\, .
 \end{equation}
Transparent sources are characterized by $\omega=0$. 

The complete emission function now becomes
 \begin{eqnarray}
   S(x,K) \, d^4x & \propto & {M_\perp \cosh(\eta-Y) \over
            (2\pi)^3 }
        \; \exp \left[- \frac{K \cdot u(x)}T \right]
        \; \exp \left[ - {r^2 \over 2 R^2}
            - {{(\eta- \eta_0)}^2 \over 2 (\Delta \eta)^2}
           \right]         \nonumber \\ & & \times 
          \exp \left [ - \frac{l_{\rm eff}}{\lambda}
          \right ] \;
            \frac{\tau \, d\tau}{\sqrt{2\pi(\Delta \tau)^2}}  
            \exp \left[- {(\tau-\tau_0)^2 \over 2(\Delta \tau)^2}
           \right ] \, d\eta \, r\, dr\, d\phi.
 \label{mod2}
 \end{eqnarray}
We do not specify the normalization of the emission function; in
principle it is fixed by the total number of produced particles.
For the {\em shape} of the single-particle spectra
and two-particle correlations the normalization is irrelevant.
As we will see a meaningful comparison with data is possible even
without knowing the normalization of the emission function and permits
us to exclude a large class of opaque source models.

The collective flow velocity profile $u(x)$ is parameterized as in
Paper I. We will deviate from the discussion presented there by
assuming a constant freeze-out temperature $T$.

\section{One- and two-particle spectra from opaque sources}
\label{modelstudy}

In this Section we show and discuss the results of numerical
calculations of correlation radii from the emission function
(\ref{mod2}). All calculations are based on the model-independent 
expressions which give the correlation radii in terms of space-time
variances of the emission function. For the Cartesian parametrization
of the correlator they read \cite{CSH95a,HB95,CSH95b}
 \begin{eqnarray}
   R_s^2({\bf K}) &=& \langle \tilde{y}^2 \rangle \, ,
 \label{3.1a}\\
   R_o^2({\bf K}) &=&
   \langle (\tilde{x} - \beta_\perp \tilde t)^2 \rangle \, ,
 \label{3.1b}\\
   R_l^2({\bf K}) &=&
   \langle (\tilde{z} - \beta_l \tilde t)^2 \rangle \, ,
 \label{3.1c}\\
   R_{ol}^2({\bf K}) &=&
   \langle (\tilde{x} - \beta_\perp \tilde t)
           (\tilde{z} - \beta_l \tilde t) \rangle \, ,
 \label{3.1d}
 \end{eqnarray}
where $\tilde x_\mu = x_\mu - \langle x_\mu \rangle$ and $\langle
\cdots \rangle$ denotes the space-time average taken with the emission
function (see Paper I). For the YKP parametrization radii we have
\cite{WHTW96,HTWW96a} 
 \begin{eqnarray}
 \label{3.2a}
   v({\bf K}) 
   &=& {A+B\over 2C} \left( 1 - \sqrt{1 - \left({2C\over A+B}\right)^2}
                     \right) \, ,
 \\
 \label{3.2b}
   R_\parallel^2({\bf K}) &=& B - v\, C \, ,
 \\
 \label{3.2c}
   R_0^2({\bf K}) &=& A - v\, C\, ,
 \\
 \label{3.2d}
   R_{\perp}^2({\bf K}) &=& {\langle{ \tilde{y}^2 }\rangle}\, ,
 \end{eqnarray}
where
 \begin{eqnarray}
 \label{3.3a}
   A &=&  \langle \tilde t^2 \rangle - \frac2{\beta_\perp} \langle
          \tilde t \tilde x \rangle + \frac1{\beta^2_\perp}
          \langle \tilde x^2 - \tilde y^2 \rangle \, ,
 \\
 \label{3.3b}
   B &=&    \langle \tilde z^2 \rangle - 2 \frac{\beta_l}{\beta_\perp} 
        \langle \tilde z \tilde x \rangle + \frac{\beta_l^2}{\beta^2_\perp}
          \langle \tilde x^2 - \tilde y^2 \rangle \, ,
 \\
 \label{3.3c}
   C &=&  \langle \tilde z \tilde t \rangle - 
        \frac{\beta_l}{\beta_\perp} \langle
        \tilde t \tilde x \rangle -
        \frac{1}{\beta_\perp} \langle
        \tilde z \tilde x \rangle 
        + \frac{\beta_l}{\beta^2_\perp}
        \langle \tilde x^2 - \tilde y^2 \rangle \, .
 \end{eqnarray}

In the following subsection we study the basic opacity effects on the
one- and two-particle spectra. Some of these are analyzed in more
detail in subsection~\ref{moredet}. A comparison with available data 
is presented in subsection~\ref{compar}.

\subsection{Basic opacity effects}
\label{basicop}

The basic free parameter of our study is the mean free path $\lambda$.
We will scan a rather wide range of opacities $\omega=R/\lambda$ 
($\omega=0,\, 1,\, 10$). For $\eta_f$, which parametrizes the
transverse flow at freeze-out linearly according to  
 \begin{equation}
 \label{eta}
    \eta(r) = \eta_f \left( {r\over R} \right)\, ,
 \end{equation}
we will study two different values, $\eta_f=0$ (a source without
transverse flow) and $\eta_f=0.5$ (rather strong transverse flow).
All other model parameters are kept fixed in this and the following 
subsection at the values listed in Table~\ref{tab1} which are
motivated by the SPS Pb+Pb data (see Sec.~\ref{compar}).
 \begin{table}
 \caption{Model parameter values used in the calculations of 
   Sec.~\ref{basicop}.}
 \begin{center}
 \begin{tabular}{lc}
 \hline \hline 
  temperature $T$ & 120 MeV \\
  average freeze-out proper time $\tau_0$ & 8 fm/$c$ \\  
  mean proper emission duration $\Delta \tau$ & 2 fm/$c$ \\
  geometric (Gaussian) transverse radius $R$ & 6.5 fm \\
  Gaussian width of the space-time rapidity profile $\Delta \eta$ & 1.3 \\
 \hline
  pion mass $m_{\pi^\pm}$ & 139 MeV/$c^2$ \\  
 \hline  \hline
 \end{tabular}
 \end{center}
 \label{tab1}
 \end{table} 
The calculations in this and the following subsection are done for
pions and pion pairs at midrapidity ($Y_{_{\rm CM}} \equiv 
Y_{\pi \pi} - Y_{_{\rm CMS}}  = 0$). 

\begin{figure}[ht]
\begin{center}
\epsfxsize=12cm
\centerline{\epsfbox{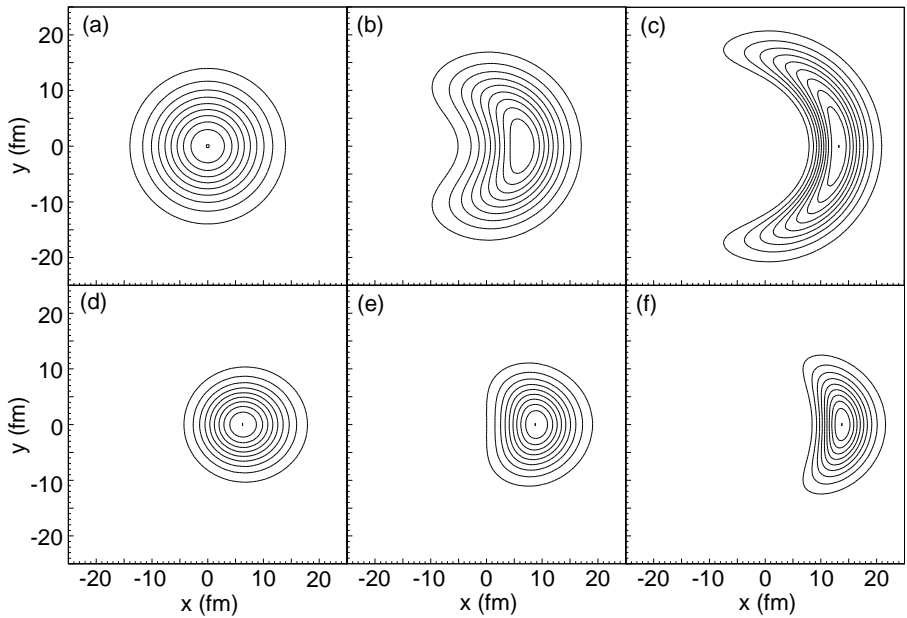}}
\caption{Transverse cuts of the emission function for midrapidity
  pions with transverse momentum $K_\perp=400$ MeV/$c$ (in
  $x$-direction), for the model parameters given in
  Table~\ref{tab1}. Top row: no transverse flow ($\eta_f = 0$),
  bottom row: transverse flow with $\eta_f = 0.5$. The columns
  correspond to different opacities: $\omega = 0$ (left), $\omega = 1$
  (middle), $\omega = 10$ (right).}
\label{f1}
\end{center}
\end{figure}

To obtain an impression of how the source changes when varying
$\omega$ and $\eta_f$ we show in Fig.~\ref{f1} for pions with
transverse momentum $K_\perp=400$ MeV/$c$ transverse cuts of the
effective emission function, $S(x,y;K_\perp=400\,{\rm MeV}/c)$, as
contour plots.
When switching on the opaqueness, the ``transverse'' size (in y- or
side-direction) of the effective source is seen to increase
dramatically (cf. Figs.~\ref{f1}a-c); this is due to the suppression 
factor $\exp[-l_{\rm eff}/\lambda]$ which cuts out all of the
interior of the source and leaves only the right hemisphere of the
dilute tail of the Gaussian transverse density distribution. This 
cannot happen in the model of Heiselberg and Vischer \cite{HV96} who use 
a transverse box profile without Gaussian tails. For a source with the 
shape given in Fig.~\ref{f1}c the Gaussian approximation, on which the
model-independent expressions from the beginning of this Section  
are based, may become questionable; for a qualitative understanding of
the relevant features it should, however, be sufficient. 

Transverse flow (lower row in Fig.~\ref{f1}) is seen to decrease the
effective source more in the sideward than in the outward direction;
this effect is the stronger the larger the opacity $\omega$.

\begin{figure}[ht]
 \begin{center}
 \epsfxsize=12cm
 \centerline{\epsfbox{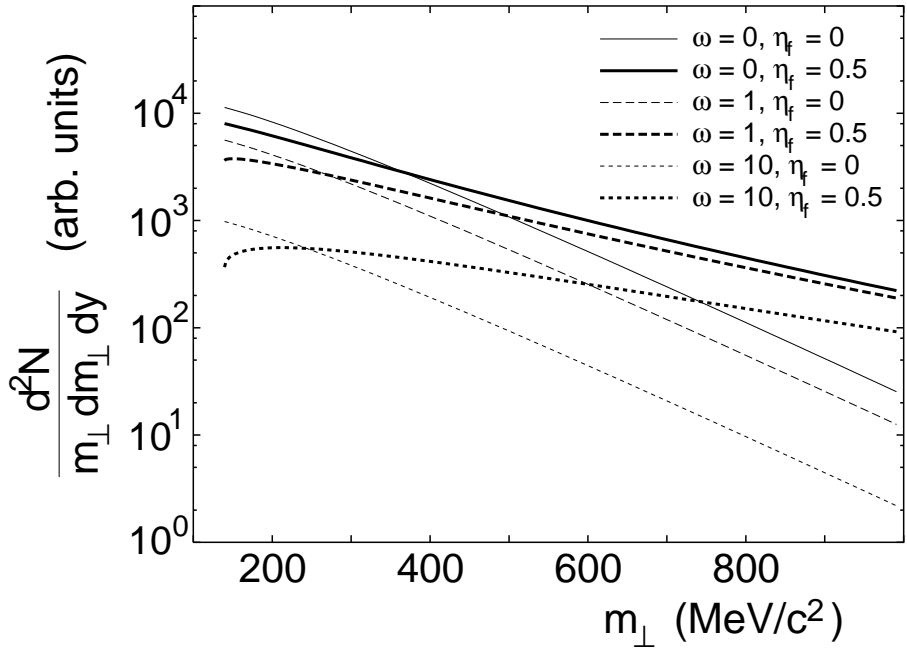}}
 \caption{Comparison of $m_\perp$ spectra for transparent and opaque sources 
  at midrapidity. For the model parameters see Table~\ref{tab1}. Thin
  lines correspond to models without transverse flow, thick lines to 
  models with $\eta_f = 0.5$. For reference, solid lines represent
  the spectra of the transparent model ($\omega = 0$). Dashed and dotted
  lines show the spectra from sources with $\omega = 1$ and $\omega = 10$
  respectively.}
 \label{f2}
 \end{center}
 \end{figure}

The single-particle transverse mass spectra resulting from these
models are shown in Fig.~\ref{f2}. 
The normalization of the spectra is arbitrary since the emission
function was not normalized. The interesting information is carried by
the spectral slopes. For non-expanding sources, the slopes for opaque
and transparent sources are identical. For expanding sources with
fixed $\eta_f$, an opaque source yields much flatter spectra than a
transparent source. The reason is obvious from Fig.~\ref{f1}:  
due to the opacity much stronger weight is given to the rapidly
expanding surface than to the less rapidly expanding interior
regions. This implies that the {\em average} flow velocity $\langle
v_\perp \rangle$ is larger for
the opaque source. In \cite{SSH93} it was argued that (for $m_\perp
\gg m_0$) the slope of the $m_\perp$-spectrum is given approximately 
by the blue-shifted temperature 
 \begin{equation}
 \label{blueshift}
   T_{\rm eff} \approx T \sqrt{\frac{1 + \langle v_\perp \rangle}
                              {1 - \langle v_\perp \rangle}} \, . 
 \end{equation}
We should therefore expect roughly the same slopes if with increasing 
opacity $\omega$ the flow parameter $\eta_f$ were reduced to keep
$\langle v_\perp \rangle$ fixed. In subsection~\ref{compar} we will
show how this works.  

\begin{figure}[ht]
 \begin{center}
 \epsfxsize=12cm
 \centerline{\epsfbox{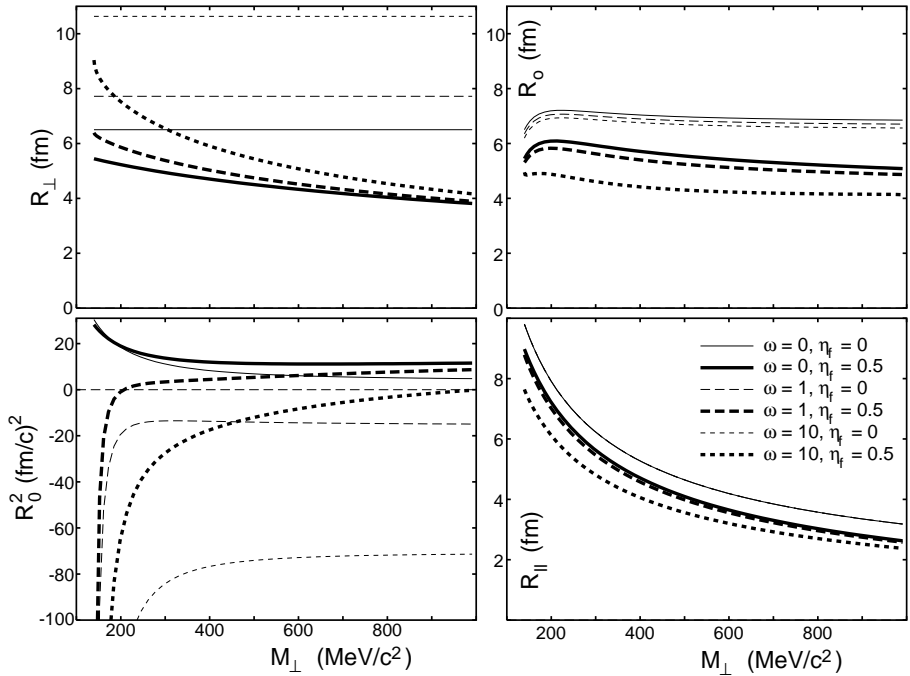}}
 \caption{Comparison of correlation radii from transparent and 
  opaque sources for pion pairs with rapidity $Y_{_{\rm CM}} = 0$.
  The source parameters are given in Table~\ref{tab1}. Thin lines:
  $\eta_f = 0$, thick lines: $\eta_f = 0.5$. Solid lines: $\omega = 0$
  (transparent source), dashed lines: $\omega =1 $, dotted lines:
  $\omega = 10$. Note that $R_s = R_\perp$ and for $Y_{_{\rm CM}} = 0 $
  the longitudinal correlation radius $R_l = R_\parallel$ and 
  $Y_{_{\rm YK}} = 0$ in LCMS.} 
 \label{f3}
 \end{center}
 \end{figure}

The two-particle correlations resulting from these models are studied
in Fig.~\ref{f3}.
For midrapidity pions the cross-term $R_{ol}$ in the Cartesian
parametrization (\ref{3.1d}) and the Yano-Koonin velocity $v$ 
of the YKP parametrization (\ref{3.2a}) vanish, and $R_l =
R_\parallel$. Furthermore, $R_s = R_\perp$ in general.

Surprisingly, without transverse flow the outward radius decreases
only weakly with increasing opacity. This is a specific feature of our
Gaussian parametrization of the source geometry and not true for a box
profile as studied by Heiselberg and Vischer \cite{HV96,HV97}. In
their case no emission from regions outside the box radius $R$ is
possible, and opacity leads to an effective emission function which is
``squeezed'' into a very thin crescent-shaped region close to the edge
of the box. In our parametrization, increasing opacity favors emission
from the more dilute tail of the Gaussian density distribution, and as
$\omega$ increases the effective emission region just moves further
outward, becoming part of a shell with a radius $R_{\rm shell} > R$
whose thickness happens to be more or less independent of $\omega$.
The sideward homogeneity radius $R_s$, on the other hand, grows
with $R_{\rm shell}$ as seen in the top row of Fig.~\ref{f1} and thus
increases dramatically with increasing opacity.

For transparent sources it is well known that transverse flow causes a
reduction of the sideward and outward homogeneity regions $\langle
\tilde y^2 \rangle$ and $\langle \tilde x^2 \rangle$, respectively,
the more so the larger $M_\perp$ \cite{CL96a,WSH96,CSH95b,CL95}. 
For fixed transverse flow parameter $\eta_f$, this effect is seen to
be even stronger for opaque sources, reflecting again the growing {\em
  average} transverse flow velocity with increasing $\omega$. 
The right upper panel of Fig.~\ref{f3} shows that for very opaque
sources this effect even eliminates the slight initial rise of $R_o$
for small $K_\perp$: the decrease of $\langle \tilde x^2 \rangle$ with
transverse mass is then stronger than the increasing contribution from
the term $\beta^2_\perp \langle \tilde t^2 \rangle$ in (\ref{3.1b}). 

It has been proposed in \cite{HV96} to identify opaque sources
by looking for a negative difference $R_o^2 - R_s^2$, especially at
$K_\perp \approx 0$. In the lower left panel of Fig.~\ref{f3} we show
that much more dramatic opacity effects are seen in the ``temporal''
YKP parameter $R_0^2$. For midrapidity pion pairs $R_0^2$ is given in
the CMS frame (which is also the rest frame of the emitting fluid element
where $v({\bf K})$ is zero) by 
 \begin{equation}
  R_0^2  = \langle {\tilde t}^2 \rangle -
  \frac {2}{\beta_\perp}  \langle \tilde x \tilde t
   \rangle + \frac 1{\beta_\perp^2}  \langle
  {\tilde x}^2 - {\tilde y}^2  \rangle \, .
 \label{3.4}
 \end{equation}
For opaque sources the last term on the r.h.s.\ is negative
and diverges to $-\infty$ as $K_\perp \to 0\; (\beta_\perp \to 0)$.
Transverse flow reduces this negative contribution, but not 
sufficiently to reverse the overall negative sign of $R_0^2$
in the region of small $K_\perp$. As seen in Fig.~\ref{f3} this
specific feature of opaque sources is quite stable against a reduction
of the opacity $\omega$ in the sense that it even shows for nearly
transparent sources with $\omega = 1$ at sufficiently large
$K_\perp$-values to be measurable.

Compared to $R_s$, $R_o$, and $R_0$, the opacity effects on
$R_l=R_\parallel$ (shown in the lower right panel of Fig.~\ref{f3})
are weak. They can all be explained in terms of the different values
for the {\em average} transverse flow velocities in the studied
examples as discussed in Paper I.

\subsection{A more detailed discussion of $R_o^2-R_s^2$ and $R_0^2$}
\label{moredet}

\begin{figure}[ht]
 \begin{center}
 \epsfxsize=12cm
 \centerline{\epsfbox{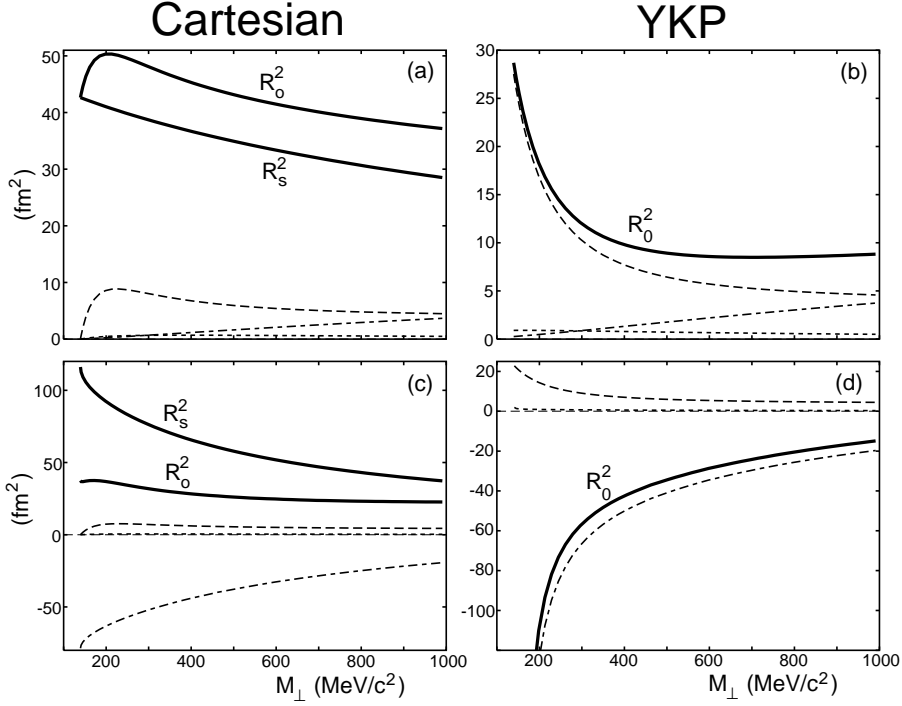}}
 \caption{Opacity effects on $R_o^2-R_s^2$ (left column) and $R_0^2$
   (right column). The solid lines show the radius parameters as they
   enter the correlation function, the other lines show how they are
   decomposed into contributions from various space-time variances.
   Left column: the dashed line shows $\beta^2_\perp \langle \tilde t^2
   \rangle$, the dotted line $-2\beta_\perp \langle \tilde x \tilde t
   \rangle$, and the dash-dotted line $\langle \tilde x^2 - \tilde y^2
   \rangle$. In the right column the same symbols denote the same
   quantities divided by $\beta_\perp^2$. All curves are calculated
   at $Y_{_{\rm CM}}=0$, with model parameters from Table~\ref{tab1}
   except for $R = 7$ fm and $\eta_f = 0.3$. Upper row: $\omega = 0$
   (transparent source); lower row: $\omega = 10$ (opaque source).}  
 \label{f4}
 \end{center}
 \end{figure}

The interesting opacity effects on $R_o^2 - R_s^2$ and on $R_0^2$ deserve 
a more detailed discussion. To this end we plot in Fig.~\ref{f4} 
the various contributions to these radius parameters from the
space-time variances of a transparent (upper row) and an opaque source
(lower row), both for the standard Cartesian (left column) and YKP
parametrizations (right column). We show results for moderate
transverse flow $\eta_f=0.3$ and midrapidity pion pairs, but we
checked that no qualitative differences occur for forward rapidities.
The diagrams in Fig.~\ref{f4} can be concisely summarized by stating
that for transparent sources the lifetime term $\langle \tilde t^2
\rangle$ dominates the difference $R_o^2 - R_s^2$ and $R_0^2$ (at
least for small $K_\perp$) while for opaque sources the difference
$\langle \tilde x^2 - \tilde y^2 \rangle$ of transverse spatial
variances takes the leading role.

These results confirm the conclusion of Ref.~\cite{HV96} that, within 
the Cartesian para\-me\-tri\-za\-tion, opaque sources generically lead to 
negative values for the difference $R_o^2-R_s^2$. Our calculations 
were done in the Yano-Koonin ($v=0$) frame\footnote{Remember that the
  Cartesian radius parameters $R_o,R_l,R_{ol}$ depend strongly on the
  longitudinal reference frame.}, and we checked that in this frame
this qualitative conclusion is independent of the pair rapidity. In
the YKP parametrization the same feature appears strongly enhanced in
the parameter $R_0^2$, especially at small values of $K_\perp$, due to
the division of the relevant combination of space-time variances by the 
factor $\beta_\perp^2$. A negative value for $R_0^2$ corresponds to a 
negative value for $R_o^2 - R_s^2$ in the Yano-Koonin frame (which is
usually not much different from the LCMS \cite{WHTW96}).

\subsection{Comparison with data}
\label{compar}

In this subsection we compare the model calculations with recently
published correlation data from 158 GeV/$c$ Pb+Pb collisions at the
SPS \cite{NA49corr}, focusing on opacity effects. To constrain the
model parameters we make use also of preliminary data on
single-particle spectra \cite{NA49QM97,NA49QM96}.     

When we increase the opacity without changing any of the other model
parameters we obtain a larger transverse HBT radius $R_\perp$ since
the opacity factor favors particle emission from the dilute tail of
the transverse Gaussian density distribution. In order to reproduce 
a fixed measured value for $R_\perp$ with sources of increasing
opacity we must therefore reduce the Gaussian width parameter $R$ in
the emission function. 

By the same mechanism larger opacities lead, at fixed $\eta_f$, to
larger average transverse flow velocities in the source. These cause a
flatter slope of the single-particle $m_\perp$-spectrum
(Fig.~\ref{f2}) and a steeper $M_\perp$-dependence of the transverse
HBT radius $R_\perp$ (Fig.~\ref{f3}). On a qualitative level both
effects can be studied analytically: for pions the inverse slope
parameter of the single-particle $m_\perp$-spectrum is given by
Eq.~(\ref{blueshift}), while the $M_\perp$-dependence of $R_\perp$ is
approximately given by \cite{CNH95,CSH95b}
 \begin{equation}
 \label{Rsapp}
   R_\perp^2 \approx 
   {R^2 \over 
   1 + \left({M_\perp\over T}\right) \langle v_\perp \rangle^2}\, .
 \end{equation}
This expression is derived from Eq.~(\ref{3.2d}) by evaluating the
average $\langle \tilde y^2 \rangle$ over the emission function via 
saddle point integration \cite{CNH95,CSH95b}. Although not
quantitatively reliable \cite{WSH96}, it illustrates nicely an important
point \cite{Sch96,H97,WTHQM97,NA49corr,NA49QM97}: while 
(\ref{blueshift}) and (\ref{Rsapp}) are, each by itself, ambiguous and
do not allow to separate $T$ and $\langle v_\perp \rangle$, {\em the
correlation between those two parameters is opposite in the two
equations}. By combining them the collective transverse flow 
$\langle v_\perp \rangle$ can be isolated from the random thermal
motion $\sim T$. The value for $T$ given in Table~\ref{tab1} stems
from a rough\footnote{In the fit resonance decay contributions
  \cite{WH97} were only included in the single-particle spectra, but
  not in the calculation of the correlation radii. Contrary to the
  analysis in \cite{NA49corr} we evaluated, however, the spectra and
  HBT radii numerically with the source (\ref{mod2}) (for $\omega=0$)
  instead of using the simple approximations (\ref{blueshift}) and
  (\ref{Rsapp}).} numerical fit \cite{WTHQM97} to the measured
single-particle spectra and correlation radii from Pb+Pb collisions at
the SPS \cite{NA49corr,NA49QM96}. We expect it to be accurate to about
$\pm 10$ MeV.  

Once $T$ is known, the average transverse flow is fixed by the slope
of the single-pion transverse mass spectrum. For $T=120$ MeV the NA49 
spectra \cite{NA49QM96} yield an average transverse flow velocity
$\langle v_\perp \rangle \simeq 0.4c - 0.45c$ \cite{BK96,H97,WTHQM97}. 
For our source model the average transverse flow velocity is given by
($\xi = r/R$) 
 \begin{eqnarray}
 \label{3.5}
 \langle v_\perp \rangle 
    & = & {1\over N R^2} \int_0^\infty r \, dr \int_0^{2\pi} d\phi \,
        v_\perp(r) \, \exp \left ( -\frac{r^2}{2\, R^2} \right ) \,
        \exp \left ( - \frac{l_{\rm eff}(r,\phi)}{\lambda} \right )\\
 \nonumber & = & \frac1N \int_0^\infty \xi\, d\xi
        \Biggl[\tanh\left(\eta_f\xi \right ) \, e^{-{\xi^2\over2}}\, 
        \int_0^{2\pi} d\phi \, \exp \left ( - \omega \,
        e^{ - {\xi^2\sin^2\phi\over 2}} \int_{\xi\cos\phi}^\infty 
        e^{ - {\zeta^2\over 2}} d\zeta \right )
        \Biggr] \, ;\\
 \nonumber
 N & = & \int_0^\infty \xi\, d\xi \left [
         e^{-{\xi^2\over2}} \int_0^{2 \pi} d\phi \, 
         \exp \left ( - \omega\, e^{-{\xi^2\sin^2\phi\over2}}\, 
         \int_{\xi\cos\phi}^\infty e^{-{\zeta^2\over2}} d\zeta \right )
         \right ] \, .
 \end{eqnarray}
This equation implies that, for fixed $\langle v_\perp\rangle$,
$\eta_f$ must be reduced if $\omega$ is increased. Therefore, if we
want to keep the absolute value of $R_\perp$, its $M_\perp$ slope, and 
the slope of the single-particle $m_\perp$-spectrum fixed, we must
decrease both $R$ and $\eta_f$ as we increase the opacity $\omega$. 

Three models with opacities $\omega = 0, \, 1,\, 10$ were
studied and compared with the data. In each case we tuned $\eta_f$ 
such that $\langle v_\perp \rangle = 0.44$ was kept fixed, and we
adjusted $R$ such that the size of $R_\perp(M_\perp)$ is roughly 
reproduced. This choice was motivated by the NA49 data shown in the 
right column of Fig.~4 of Ref.~\cite{NA49corr} (HBT radii from 
158 GeV/$c$ Pb+Pb collisions, $h^-h^-$ correlations in the rapidity window 
$3.9 < Y_{\pi\pi} < 4.4$). The values of $R$ and $\eta_f$ used in these 
calculations are listed in Table~\ref{tab2}. 
 \begin{table}
  \caption{Transverse flow strength $\eta_f$ and Gaussian radius $R$
  for the three models which are used in the comparison with the data
  in Sec.~\ref{compar}.}
  \begin{center}
  \begin{tabular}{lcc}
  \hline \hline
  $\omega$ (opacity) & $\eta_f$ & $R$ \\
  \hline
   0 (transparent) & 0.4 & 5.74 fm \\
   1 & 0.345 & 4.83 fm \\
   10 (opaque) & 0.215 & 3.35 fm \\
   \hline \hline
  \end{tabular}
  \end{center}
  \label{tab2}
 \end{table}
To fit the parallel correlation radius $R_\parallel$ (see later) the 
average freeze-out time $\tau_0$ has been set to 6.1~fm. Other model 
parameters are taken from Table~\ref{tab1}.

\begin{figure}[ht]
 \begin{center}
 \epsfxsize=12cm
 \centerline{\epsfbox{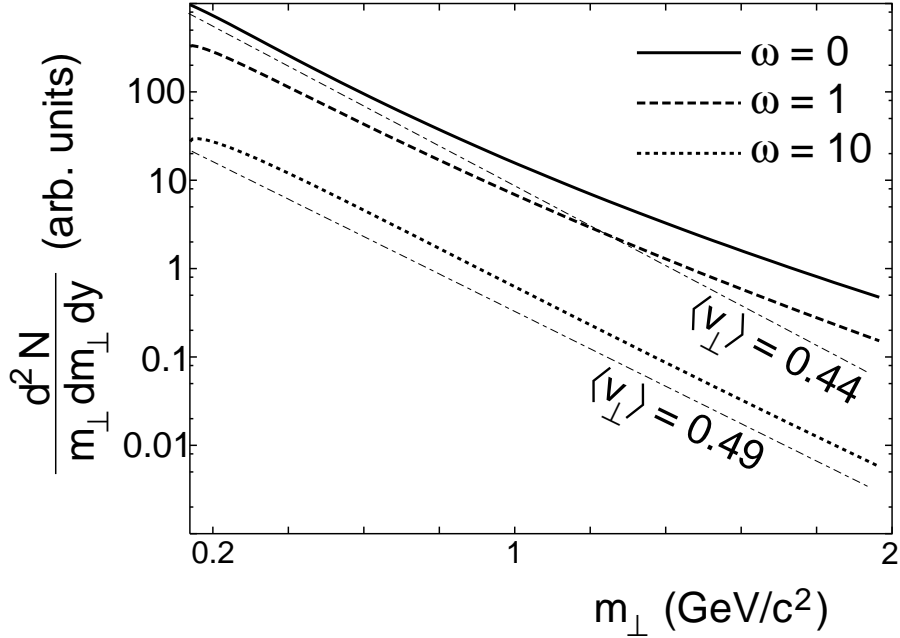}}
 \caption{Single-particle $m_\perp$-spectra from the models listed in 
   Table~\ref{tab2}. Solid, dashed, and dotted  lines correspond to 
   models with $\omega = 0,\, 1,\, 10$, respectively. The thin  
   dash-dotted auxiliary lines show a straight exponential with
   inverse slope $T_{\rm eff} = T \sqrt{(1+\langle v_\perp \rangle) /
   (1-\langle v_\perp \rangle)}$, for $T=120$ MeV and $\langle v_\perp
   \rangle=0.44$ resp.\ $\langle v_\perp \rangle = 0.49$. 
   The spectra are computed at midrapidity where the
   analysis of \cite{BK96} was performed.}
 \label{f5}
 \end{center}
 \end{figure}

In Fig.~\ref{f5}
we plot the single-particle $m_\perp$-spectra resulting from these 
models. (Recall that only the slopes, but not the normalization are
relevant here.) According to Eq.~(\ref{blueshift}) we expect the same
slopes for all spectra, which should be characterized by the blue-shifted
temperature with $\langle v_\perp \rangle = 0.44$.
This is roughly borne out by the calculation, but for transparent and
mildly opaque sources the spectra are visibly concave \cite{SSH93} and 
show an increasingly flatter slope for $m_\perp > 1$ GeV. The reason
for this is well-known \cite{SSH93}: for large $m_\perp$ the spectra
are dominated not by particles from fluid cells with the average
transverse flow velocity $\langle v_\perp \rangle$, but from more
rapidly moving fluid cells in the tail of the transverse density
profile. This detail cannot be accurately accounted for by the
simple formula (\ref{blueshift}). In other words: for a specified 
$p_\perp$ there is only a part of the source contributing to the
particle production -- this is the effective source for that
$p_\perp$. The slope of the spectra at that value of $p_\perp$
is given by the blue-shifted temperature where the average transverse
expansion velocity is to be calculated over the {\em effective} source.
For the very opaque source this average velocity is (accidentally) 
independent of $p_\perp$, and the spectrum is thus exactly exponential. 
However, to reproduce its slope via (\ref{blueshift}) we need 
$\langle v_\perp \rangle = 0.49$, a somewhat larger value than the 
average transverse expansion velocity computed over the complete source.
Note that the experimental spectra resemble more the concave spectrum 
of the transparent source than the straight line originating from 
the opaque one.

\begin{figure}[ht]
 \begin{center}
 \epsfysize=11cm
 \centerline{\epsfbox{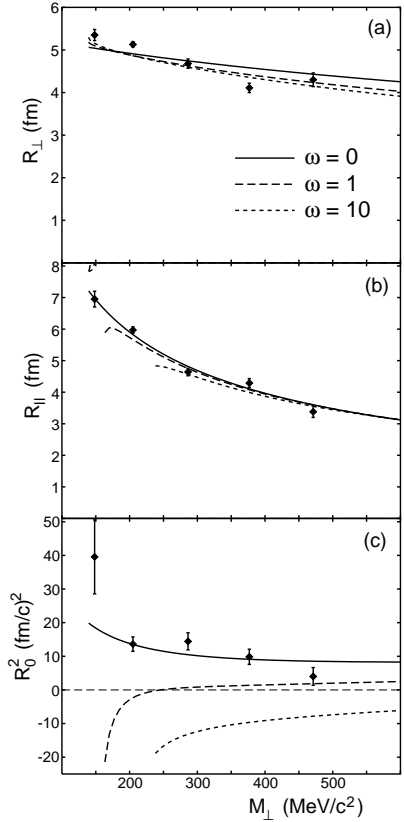}}
 \caption{$M_\perp$ dependences of the YKP correlation radii at
   $Y_{_{\rm CM}} = 1.25$ resulting from the three models
   in Table~\ref{tab2}. (Same line symbols as in Fig.~\ref{f5}.) 
   The data are from 158 $A$ GeV/$c$ Pb+Pb collisions taken by the NA49
   collaboration \cite{NA49corr}.}
 \label{f6}
 \end{center}
 \end{figure}

Fig.~\ref{f6} shows
the corresponding set of YKP correlation radii, evaluated at 
$Y_{_{\rm CM}} = 1.25$ where the NA49 collaboration has published the 
$M_\perp$ dependence of the three YKP radius parameters from 158 $A$ 
GeV/$c$ Pb+Pb collisions at the SPS (Fig.~4 in Ref.~\cite{NA49corr}). 
Note that the systematic error of the data, estimated to be $\pm 15\%$
\cite{NA49corr}, is not included in the plots. In Fig.~\ref{f6}a one 
sees that by simply adjusting the source parameters to keep the magnitudes
of $R_\perp$ and $\langle v_\perp \rangle$ approximately constant, the 
$M_\perp$-dependence of $R_\perp$ is reasonably well reproduced by all 
three models. This implies that the average transverse flow velocity is 
indeed the dominating factor for the $M_\perp$-slope of the transverse 
HBT radius. We found that the agreement between model and data can be 
further improved by additional fine-tuning of the parameters, in particular
by taking into account a non-vanishing transverse temperature gradient. 
Since a fully quantitative analysis requires, however, also the inclusion 
of resonance decays which may cause an additional $M_\perp$ dependence 
\cite{WH97,SX96}, and since the following arguments will not depend on 
such details we will not enter into this discussion here.
 
The longitudinal radius parameter $R_\parallel$ shows even less
sensitivity to the opacity $\omega$ (Fig.~\ref{f6}b). The reason is
that it is determined mainly by the values $\tau_0$ (average
freeze-out proper time) and $\Delta \tau$ (mean emission duration), 
and its $M_\perp$ dependence is dominated by the boost-invariant 
longitudinal expansion \cite{MS88,WHTW96}. Since, except for very 
forward and backward rapidities, opacity affects the emission function 
mostly in the transverse direction, $R_\parallel$ is hardly modified. 
By appropriately adjusting $\tau_0$, the data for $R_\parallel$ are very 
well reproduced. Note that for opaque sources the calculated curves 
feature a gap at small values of $M_\perp$ and behave strangely near 
$M_\perp = m_\pi$; the origin of this behaviour will be explained
in Sec.~\ref{modykp}. This problem does not arise at midrapidity where
unpublished data on the $M_\perp$-dependence of $R_\parallel$ can be 
found in the PhD theses \cite{appelsPhD} which 
we were able to fit equally well all the way down to $K_\perp=0$ 
with our emission function for all three values of $\omega$, using 
$\tau_0=8$ fm/$c$.

Since the space-time rapidity width $\Delta\eta=1.3$ is fixed by the
width of the single-particle rapidity distribution
\cite{Sch96,NA49corr} and $\tau_0$ has been set to the value of 6.3~fm, 
at this point the emission duration $\Delta \tau$ is the only remaining 
free parameter in the emission function. For transparent sources 
($\omega=0$) it should be fixed by the asymptotic value of $R_0$ at 
large values of $M_\perp$ \cite{HTWW96a}. (Its effects on $R_\perp$ 
are negligible, and for $\Delta\tau \ll \tau_0$ they are also small 
for $R_\parallel$ \cite{WHTW96}.) In Fig.~\ref{f6}c we compare with the
data our results for $R_0^2$ for a fixed value $\Delta\tau = 2$ fm/$c$, 
but for three different opacities. Clear disagreement with the data is 
seen for the opaque models. Already the still rather transparent model 
with $\omega = 1$ (cf. Fig.~\ref{f1}) misses nearly all data points.

The question arises to which extent this disagreement can be avoided
by readjusting the duration of particle emission $\Delta\tau$. As
mentioned before and seen in Fig.~\ref{f6}c, for opaque sources
the asymptotic value of $R_0^2$ at large $M_\perp$ is no longer
given by the effective source lifetime $\langle \tilde t^2\rangle$ in
the YK frame alone, but also receives a strong negative geometric
contribution $\langle \tilde x^2 - \tilde y^2 \rangle$ (cf. 
Eq.~(\ref{3.4})). One might try to compensate this effect by increasing 
for opaque sources the lifetime $\Delta\tau$. This is physically not
unreasonable because opaque sources emit particles only from the surface;
to produce the same total particle yield the resulting reduced 
brightness must be compensated by a larger emission duration. However,
larger values of $\Delta\tau$ lead also to an increase of $R_\parallel$
which must be compensated by reducing the average freeze-out time 
$\tau_0$. Since our model parametrization looses its meaning for 
$\Delta\tau > \tau_0$, there are clear limits to what can be achieved 
in this way. 

\begin{figure}[ht]
 \begin{center}
 \epsfysize=11cm
 \centerline{\epsfbox{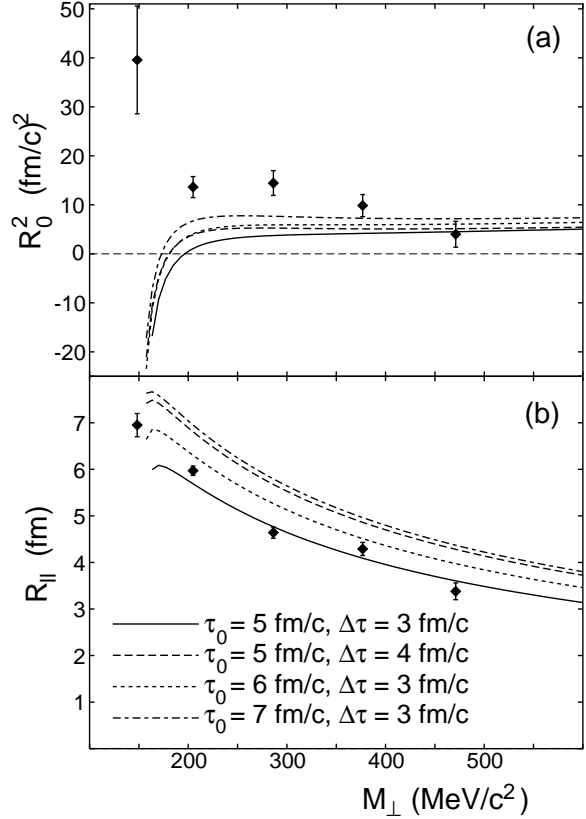}}
 \caption{The YKP radius parameters $R_0^2$ (a) and $R_\parallel$ (b)
  computed for $\omega = 1$ and different values of $\tau_0$ and 
  $\Delta \tau$ at $Y_{_{\rm CM}} = 1.25$. Solid lines: 
  $\tau_0 = 5 \, \mbox{fm/}c, \, \Delta \tau = 3 \, \mbox{fm/}c$;
  dashed lines: $\tau_0 = 5 \, \mbox{fm/}c, \, \Delta \tau = 4 \, 
  \mbox{fm/}c$; dotted lines: $\tau_0 = 6 \, \mbox{fm/}c, \, 
   \Delta \tau = 3 \, \mbox{fm/}c$; dash-dotted lines:
  $\tau_0 = 7 \, \mbox{fm/}c, \, \Delta \tau = 3 \, \mbox{fm/}c$.
  Other model parameters are taken from the corresponding calculation
  in Fig.~\ref{f6}.}
 \label{f6b}
 \end{center}
 \end{figure}

In Fig.~\ref{f6b}
we show $R_\parallel$ and $R_0^2$ for a mildly opaque source with
$\omega = 1$ for different combinations of $\tau_0$ and $\Delta\tau$
and compare them with the data \cite{NA49corr}. For $R$ and $\eta_f$ 
we took the corresponding values from Table~\ref{tab2}.

One sees that the largest lifetime $\Delta \tau$ compatible with 
the large-$K_\perp$ data for $R_\parallel$ is 4 fm/$c$. To maintain 
a reasonable description of $R_\parallel$ in this case $\tau_0$ must 
be reduced to 5~fm/$c$ which brings it dangerously close to 
$\Delta\tau$. Still, the corresponding curves for $R_0^2$ all miss 
the first data points by a large margin. Since however, as already seen 
in Fig.~\ref{f6}, at the given value of the pair rapidity the calculated 
curves for $R_0^2$ feature a gap at low $K_\perp$ (indicating that
for our model the YKP parametrization is not well-defined in this
momentum region, see Sec.~\ref{modykp}), one may question the significance
of this failure of our model to reproduce the data. For this
reason we also checked the behaviour of $R_0^2$ and $R_\parallel$ in 
different rapidity bins, using the unpublished data of 
Ref.~\cite{appelsPhD}. In all cases the data points for $R_0^2$ 
at the lowest $K_\perp$ values are positive, while the calculated curves
for opaque sources with $\omega \geq 1$ give strongly negative values, 
missing the data by far. In particular, this is true at midrapidity where
the YKP parametrization remains well-defined down to $K_\perp=0$ even for
strongly opaque sources.

We conclude that the NA49 data appear to exclude source
opacities $\omega > 1$, i.e. sources with a ``surface thickness''
$\lambda < R$. The pion source created in Pb+Pb collisions at the CERN 
SPS seems to be rather ``transparent'' at freeze-out, i.e. at the
end of the hydrodynamical expansion phase the pions freeze-out in
bulk, decoupling essentially everywhere at the same time, including 
the center of the collision region. It must be kept in mind, however, 
that this conclusion rests heavily on the positive experimental values 
for $R_0^2$ at small $K_\perp$ reported by the NA49 collaboration.
In \cite{HV96,HV97} it was remarked that preliminary NA44 data from 
Pb+Pb collisions at the SPS indicate that at small $M_\perp$ in the 
Cartesian parametrization $R_o^2$ might be smaller than $R_s^2$ (see 
\cite{NA44QM97}); this would favor an opaque source. In our opinion it 
is premature to draw firm conclusions from these data because, unlike 
NA49 \cite{NA49corr}, the NA44 Collaboration has not yet presented a 
YKP fit of their data, where the opacity signal should be much clearer.  
Taking their Cartesian fit presented in \cite{NA44QM97} at face value
we conclude that, while a mildly opaque source with $\omega \simeq 1$
cannot be excluded, the NA44 data are also inconsistent with strongly
surface dominated emission ($\omega > 10$).

In the next Section we will explain the origin of the missing pieces 
in the curves for $\omega>1$ in Figs.~\ref{f6} and \ref{f6b}. 

\section{Modified YKP parametrization}
\label{modykp}

In the previous Section we discovered an intrinsic
deficiency of the YKP parametrization which had gone unnoticed 
before. In this Section we explain the limitations of the YKP
parametrization and suggest a slightly different parametrization
which avoids these problems and should be particularly suitable for
strongly opaque, azimuthally symmetric sources with dominant
longitudinal expansion.  

\subsection{Forward rapidity pion pairs from opaque sources}
\label{opforw}

\begin{figure}[ht]
 \begin{center}
 \epsfysize=11cm
 \centerline{\epsfbox{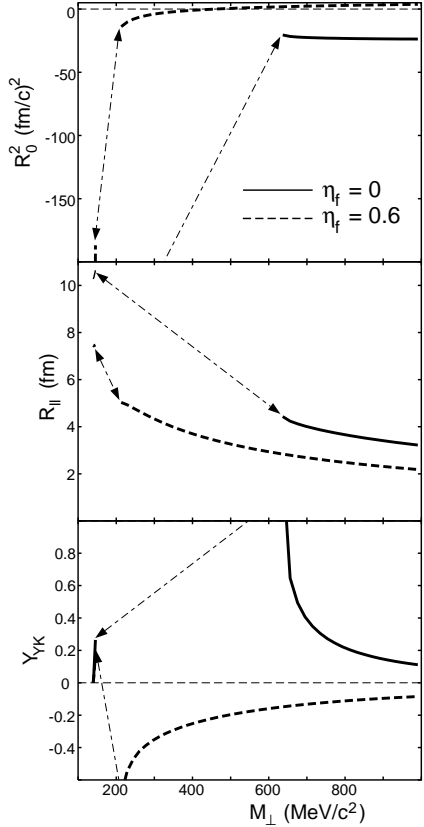}}
 \caption{The YKP radii and the Yano-Koonin rapidity 
   in the LCMS for pion pairs with $Y_{_{\rm CM}} = 1.5$
   from an opaque source ($\omega = 10$) with $R=4$ fm. Solid lines:
   $\eta_f = 0$. Dashed lines: $\eta_f = 0.6$. The other model
   parameters are taken from Table~\ref{tab1}. The arrows connect the
   curves across the gap in which the YKP parametrization is not
   defined.} 
 \label{f7}
 \end{center}
 \end{figure}

In Fig.~\ref{f7}
we show the YKP radius parameters $R_0^2$ and $R_\parallel$ and 
the YK rapidity $Y_{_{\rm YK}} = {1\over 2} \ln [(1+v)/(1-v)]$
(relative to the LCMS) for pion pairs at forward rapidity 
$Y_{_{\rm CM}} =1.5$ for opaque sources with and without transverse
flow. Let us first discuss the Yano-Koonin rapidity $Y_{_{\rm YK}}$:
For transparent sources we know from \cite{WHTW96,HTWW96a} that the YK
rapidity is negative in the LCMS, i.e. that the effective source moves
somewhat more slowly in the beam direction than the emitted pairs. For
opaque sources this seems no longer to be true; for example,
in Fig.~\ref{f7} in the absence of transverse flow the YK rapidity
comes out positive in the LCMS. We hasten to stress that this does
{\em not} imply that in this case the effective source moves faster
than the emitted pairs (which would indeed be counterintuitive); it
rather reflects the fact that for opaque sources the geometric
correction terms in the expression for the YK velocity (see Eq.~(4.1)
in Ref.~\cite{WHTW96}), in particular the one proportional to $\langle 
\tilde x^2 - \tilde y^2 \rangle$, are large and {\em spoil the 
interpretation of $v$ as the longitudinal source velocity}.

Figs.~\ref{f6} and \ref{f7} show that these correction terms can cause
even more severe problems: in certain $M_\perp$-regions the argument
of the square root in (\ref{3.2a}) turns negative and the YK velocity 
(and thus also $R_0^2$ and $R_\parallel^2$, see (\ref{3.2b}),
(\ref{3.2c})) becomes undefined. The conditions for the occurrence of
such a pathological behaviour are discussed in Appendix \ref{AS1}. It
appears that it is connected in a generic way to $\langle \tilde x^2
\rangle < \langle \tilde y^2 \rangle$ which, as we have seen,
characterizes opaque sources. 

Note that the mentioned ``strange'' behaviour of the YK velocity, in the
regions where it is defined,  is reflected via 
Eqs.~(\ref{3.2b}) and (\ref{3.2c}) in also unexpected behaviour of 
$R_\parallel$ and $R_0^2$ especially near to the gap.

\subsection{A solution of the problem}
\label{mykp}

This problem of the YKP parametrization can be avoided by a simple
technical modification which we call ``Modified
Yano-Koonin-Podgoretski\u\i '' parametrization. It differs from the 
YKP parametrization by using $q_s$ rather than $q_\perp = \sqrt{q_s^2
  + q_o^2}$ as one of the three independent components of the relative
momentum $q$. For distinction we denote the corresponding HBT
parameters with a prime: 
 \begin{eqnarray}
 \label{5.3}
   C({\bf q},{\bf K}) = 1 + \exp\Bigl[
   - R_\perp^{\prime 2}({\bf K})\, q_{s}^2 
   - R_\parallel^{\prime 2}({\bf K}) \left( q_l^2 - (q^0)^2 \right) \\
\nonumber
   - \left( R_0^{\prime 2}({\bf K}) + R_\parallel^{\prime 2}({\bf K})\right)
         \bigl(q\cdot U'({\bf K})\bigr)^2 \Bigr] 
 \end{eqnarray}
with
 \begin{equation}
 \label{5.3a}
   U'({\bf K}) = \gamma'({\bf K}) (1,0,0,v'({\bf K}))\, .
 \end{equation}
Introducing in analogy to (\ref{3.3a})-(\ref{3.3c}) the shorthands
 \begin{eqnarray}
 A' & = & \langle \tilde t^2 \rangle - \frac2{\beta_\perp} \langle
          \tilde t \tilde x \rangle + \frac1{\beta^2_\perp}
          \langle \tilde x^2 \rangle \, ,
 \label{5.4} \\   
 B' & = & \langle \tilde z^2 \rangle - 2 \frac{\beta_l}{\beta_\perp} 
        \langle \tilde z \tilde x \rangle + \frac{\beta_l^2}{\beta^2_\perp}
          \langle \tilde x^2 \rangle \, ,
 \label{5.5} \\  
 C' & = & \langle \tilde z \tilde t \rangle - 
        \frac{\beta_l}{\beta_\perp} \langle
        \tilde t \tilde x \rangle -
        \frac{1}{\beta_\perp} \langle
        \tilde z \tilde x \rangle 
        + \frac{\beta_l}{\beta^2_\perp}
        \langle \tilde x^2 \rangle \, ,
 \label{5.6}     
 \end{eqnarray}
the modified YKP parameters can be expressed by the same formulae as
the original ones:
\begin{eqnarray}
 \label{5.7}
   v' &=& {A'+B'\over 2C'} 
          \left( 1 - \sqrt{1 - \left({2C'\over A'+B'}\right)^2} \right) \, ,
 \\
 \label{5.8}
   R_\parallel^{\prime 2} &=& B' - v' C' \, ,\\
 \label{5.9}
   R_0^{\prime 2} &=& A' - v' C' \, ,\\
 \label{5.10}
   R_\perp^{\prime 2} &=& \langle \tilde y^2 \rangle\, .
\end{eqnarray}
In the modified YK frame, defined by $v'=0$, we thus have
$R_\parallel^{\prime 2} =B'$ and $R_0^{\prime 2}=A'$. Of course, 
since both YKP and Modified YKP parametrizations are just two ways of 
parameterizing the same correlation function, they are related in a simple 
way:
\begin{eqnarray}
\label{relation1}
A & = & \gamma^{\prime 2} R_0^{\prime 2} 
        + \gamma^{\prime 2}v^{\prime 2} R_\parallel^{\prime 2} -
        \frac 1{\beta_\perp^2} R_\perp^{\prime 2} \, , \\
\label{relation2}
B & = & \gamma^{\prime 2} R_\parallel^{\prime 2} 
        + \gamma^{\prime 2}v^{\prime 2}R_0^{\prime 2} -
        \frac {\beta_l^2}{\beta_\perp^2} R_\perp^{\prime 2} \, ,\\
\label{relation3}
C & = & (R_0^{\prime 2} + R_\parallel^{\prime 2}) \gamma^{\prime 2} v' - 
        \frac{\beta_l}{\beta_\perp^2} R_\perp^{\prime 2} \, .
\end{eqnarray}
The inverse relations are given by
\begin{eqnarray}
\label{inrelation1}
A' & = & \gamma^2 R_0^2 + \gamma^2 v^2 R_\parallel^2 
        + \frac1{\beta_\perp^2} R_\perp^2 \, , \\
\label{inrelation2}
B' & = & \gamma^2 R_\parallel^2 + \gamma^2 v^2 R_0^2 + 
        \frac{\beta_l^2}{\beta_\perp^2} R_\perp^2 \, , \\
\label{inrelation3}
C' & = & (R_0^2 + R_\parallel^2) \gamma^2 v + 
        \frac{\beta_l}{\beta_\perp^2} R_\perp^2 \, .
\end{eqnarray}

In Appendix~\ref{AS1} we show that the Modified YKP parametrization is
defined everywhere except for the point $K_\perp = 0$ to which it can
be smoothly extrapolated. Furthermore, whereas the relative momentum
components used in the original YKP parametrization satisfy the inequality
 \begin{equation}
    q_\perp \ge q_o = 
    \frac1{\beta_\perp} q^0 - \frac{\beta_l}{\beta_\perp} q_l \, ,
 \label{5.11}
 \end{equation}
which means that the data points never fill the whole
three-dimensional $q$-space, no such restriction exists for $q_s$
which is used in the modified parametrization. This should help to
avoid certain technical problems in the fitting procedure which can
occur with the YKP parametrization \cite{lasiuk}. In any case, it may
be useful to check the YKP fit against a modified YKP fit via the
relations (\ref{relation1})-(\ref{relation3}), in order to avoid
the pitfalls related to the possible non-existence of a YKP
parametrization for the data under study (which may not show up
clearly in the fitting process but might cause it to converge to a
wrong result). We would strongly recommend this check in order
to support our conclusion from the previous section that the data
exclude opaque sources; this conclusion was based on the value of 
$R_0^2$ in a region where, if the source were indeed opaque, the 
existence of YKP parametrization might be questionable. 

\begin{figure}[ht]
 \begin{center}
 \epsfxsize=12cm
 \centerline{\epsfbox{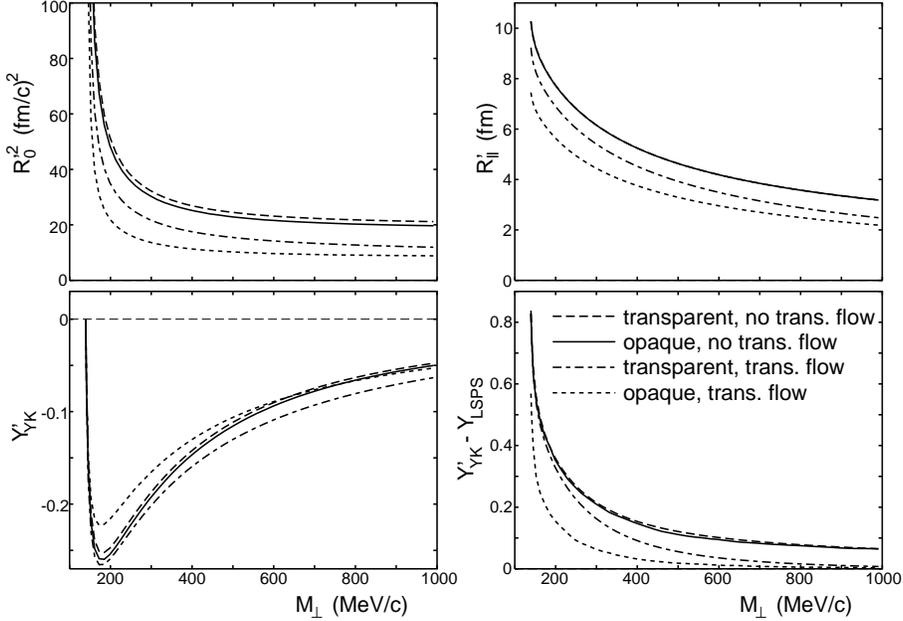}}
 \caption{The correlation radii ${R_0^{\prime}}^2$ and
   $R_\parallel^\prime$ of the Modified YKP parametrization, for pions
   at $Y_{_{\rm CM}} = 1.5$. Dashed and dash-dotted lines correspond to 
   the transparent source without transverse flow (dashed) and with
   $\eta_f = 0.6$ (dash-dotted). Solid and dotted lines correspond
   to an opaque source ($\omega = 10$) with $\eta_f = 0$ (solid)
   and $\eta_f = 0.6$ (dotted). The lower row shows the modified
   Yano-Koonin rapidity $Y_{_{\rm YK}}^{\prime}$ in the LCMS (left)
   and relative to the LSPS (right) as a function of $M_\perp$.
   Other source parameters as in Fig.~\ref{f7}.}
 \label{f8}
 \end{center}
 \end{figure}

Unfortunately, the physical interpretation of the modified YKP radii 
is no more as straightforward as that of the original ones. Wherever 
in $A,B,C$ the difference $\langle \tilde x^2 - \tilde y^2\rangle$ 
occurs it is now replaced in $A', B', C'$ by $\langle \tilde x^2 
\rangle$ alone. For transparent sources the difference $\langle \tilde 
x^2 - \tilde y^2\rangle$ is usually small \cite{WHTW96} and the 
corresponding correction terms to the leading geometric contributions 
to $R_0^2$ and $R_\parallel^2$ are of minor importance; this permits a 
direct interpretation of $R_0$ and $R_\parallel$ as the effective 
lifetime and longitudinal size of the source in its own rest frame 
\cite{HTWW96a}. For such sources the occurrence of $\langle \tilde x^2 
\rangle$ alone in the modified radius parameters is certainly a 
drawback and usually leads to large corrections which invalidate a 
naive geometrical space-time interpretation (see Fig.~\ref{f8}).  
For opaque sources the appearance of $\langle \tilde x^2 \rangle$ 
(which is related to the curvature and thickness of the emitting 
surface shell) instead of the generically much larger combination 
$\langle \tilde x^2 - \tilde y^2\rangle$ may at first sight appear as 
an advantage. Nevertheless, unless the emitting surface layer is 
indeed {\em very} thin (and very flat!), the resulting correction term 
in particular to the leading contribution in $R'_0$ cannot be 
neglected and spoils its simple interpretation as an effective source 
lifetime. This is shown in the left upper panel of Fig.~\ref{f8}; 
even for opaque sources with $\omega = 10$, one sees that $R'_0$ is in
most of the cases much bigger than the effective source lifetime
$\Delta \tau = 2$ fm/$c$, even at large values of $M_\perp$ (where
$\beta_\perp \approx 1$).  

\begin{figure}[ht]
 \begin{center}
 \epsfxsize=12cm
 \centerline{\epsfbox{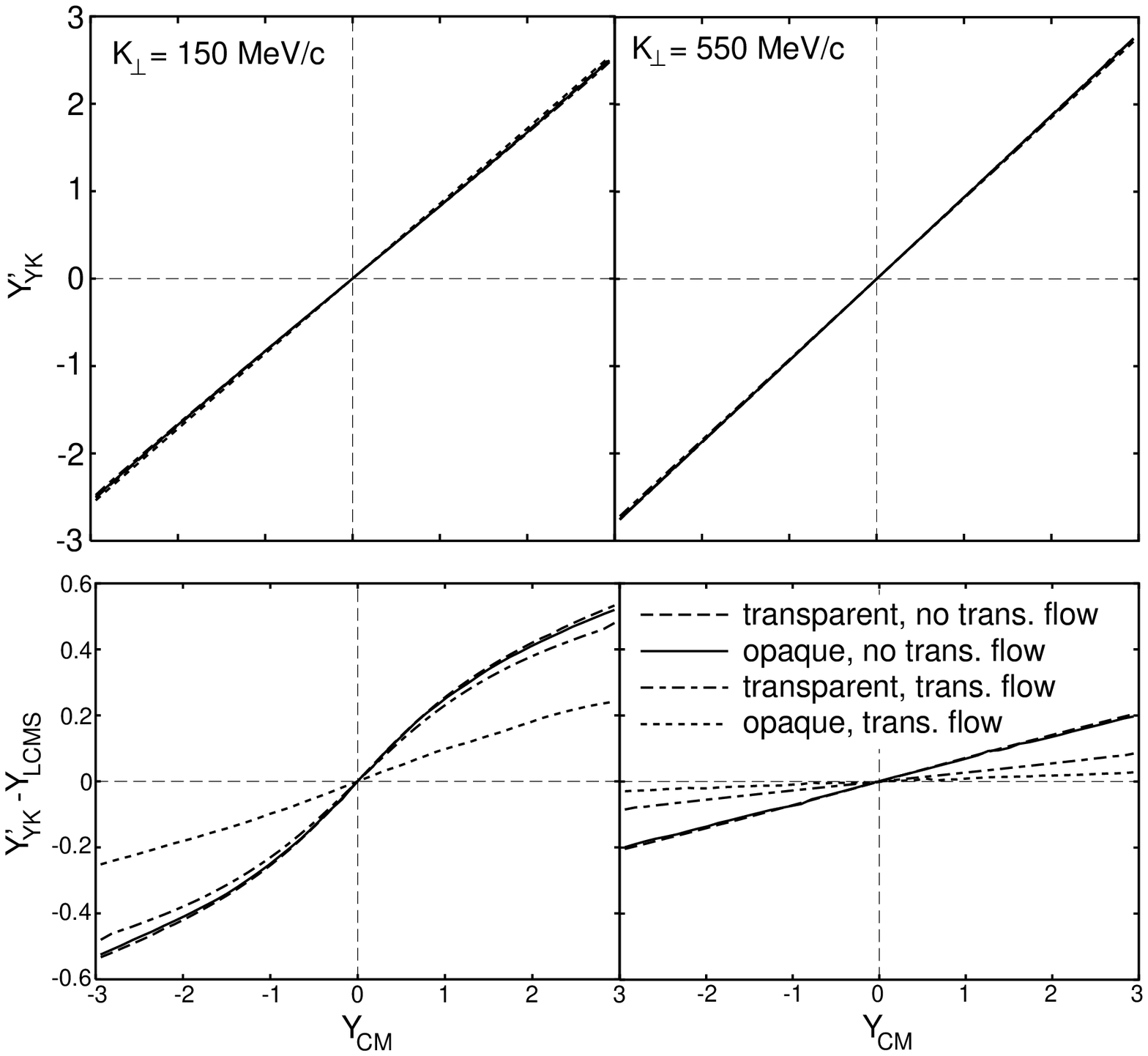}}
 \caption{The modified YK rapidity $Y_{_{\rm YK}}^{\prime}$ in the CMS
   (upper row) as a function of the pion pair rapidity in the CMS,
   for two values of $K_\perp = 150 \, \mbox{\rm MeV/}c$ (left)
   and 550~MeV/$c$ (right). Line symbols and model parameters
   as in Fig.~\ref{f8}.}
 \label{f9}
 \end{center}
 \end{figure}

The effects on the modified longitudinal radius parameter 
$R'_\parallel$ (right upper panel in Fig.~\ref{f8}) are much smaller, 
due to the small value of the longitudinal pair velocity $\beta_l$ in 
the modified YK frame (where $v'=0$) which multiplies the correction 
terms. $\beta_l$ is small in the modified YK frame because, like the 
original YK rapidity, $Y'_{_{\rm YK}}$ rises linearly with the pair
rapidity $Y_{_{\rm CM}}$ with nearly unit slope, reflecting the
boost-invariant longitudinal expansion of the source (see upper row of
Fig.~\ref{f9}).   
The difference between $Y'_{_{\rm YK}}$ and the longitudinal flow rapidity 
$Y_{_{\rm LSPS}}$ around the point of maximum emissivity (see lower 
row of Fig.~\ref{f9}) is somewhat larger than for the original YK 
rapidity for transparent sources \cite{HTWW96a,WHTW96}, but still small 
enough to consider $Y'_{_{\rm YK}}$ as a good approximation for the 
rapidity of the effective source. The difference between $Y'_{_{\rm 
YK}}$ and the rapidities of the LCMS and LSPS disappears in the limit 
$K_\perp \to \infty$, as for the original YK rapidity. Opacity effects 
on $Y'_{_{\rm YK}}$ are seen to be small.

\section{Conclusions}
\label{conclusions}

The most important result found here is that a significant opaqueness
of the source leads to dramatic effects on the YKP radius parameter
$R_0^2$. Opaqueness was parametrized by the opacity, i.e. the ratio
between the Gaussian transverse geometric radius $R$ and the surface 
thickness or mean free path, $\omega = R/\lambda$. Even for moderate
values of $\omega$, $\omega \sim 1$, $R_0^2$ becomes negative
at small, but experimentally easily accessible values of
$K_\perp$. For higher opacities this feature extends over larger
$M_\perp$ regions. For $\omega > 1$ the effect should be clearly
visible in the existing data from Pb+Pb collisions at CERN. 

However, we also encountered a problem which made the comparison
with data slightly problematic: we discovered previously unknown regions 
of inapplicability of the YKP parametrization of the correlation function.
These regions of inapplicability are connected with the use of 
$q_\perp= \sqrt{q_s^2+q_o^2}$ instead of $q_s$ as an independent relative 
momentum variable. We found that for opaque sources the YK velocity may 
become ill-defined. For this situation we suggested a modified YKP 
parametrization which is always well-defined and particularly suited 
for opaque sources. Even for transparent sources it can always be used 
as a technical tool to check the correct convergence of the
YKP fit, by using the relations (\ref{relation1})-(\ref{relation3}).
On the physical level, the modified YK velocity $v'$ can again be
interpreted, in good approximation, as the longitudinal velocity of
the effective source. It continues to reflect the longitudinal
expansion of the source through a strong dependence on the rapidity of
the emitted pairs. The interpretation of the modified transverse and
longitudinal radius parameters $R'_\perp$ and $R'_\parallel$ as
transverse and longitudinal regions of homogeneity in the source rest
frame remains valid with sufficient approximation. The interpretation
of $R_0^{\prime 2}$ (which is now always positive) as the square of
the effective source lifetime, however, is spoiled by a large
geometric correction. The extraction of a reliable estimate of the
source lifetime for opaque sources thus appears to be a very difficult
problem.   

From the fact that the published data \cite{NA49corr} show
only positive values for $R_0^2$ we concluded that the experiment 
favors a source with volume-dominated emission. Pion freeze-out in
heavy-ion collisions thus appears to be similar to the decoupling of
the microwave background radiation in the Early Universe: at a certain
point in time, when the matter has become sufficiently dilute and
cool, suddenly the entire fireball (universe) becomes transparent.
However, since for large opacities and in the rapidity region covered
by the published data \cite{NA49corr} the YKP parametrization was
found to break down in the critical region of negative $R_0^2$, two
additional checks are required to confirm this conclusion: the
$M_\perp$-dependence of the YKP radii at midrapidity should be
published, and a careful cross-check of the fitted correlation radii
with the Cartesian and/or modified YKP parametrizations should be
performed, using the relations given in Sec.~\ref{modykp} and in
\cite{WHTW96,HTWW96a}. In addition to boosting our confidence in the
correct convergence of the multidimensional fit to the measured
correlation function, such a cross-check would also exclude possible
doubts about the applicability of the YKP parametrization to the
data. Without doing that one can still try to draw conclusions based
on the difference $R_o^2 - R_s^2$ at small $K_\perp$, but compared to
the opacity signal carried by $R_0^2$ this difference is expected to
be smaller and more difficult to measure accurately. 

\vspace{4ex}

{\bf Acknowledgments:} We are indebted to Harry Appelsh\"auser,
Daniel Ferenc, Henning Heiselberg, Axel Vischer, and Urs Wiedemann for
stimulating and clarifying discussions. U.H. thanks the Institute for
Nuclear Theory in Seattle for its hospitality and for providing a
stimulating environment while this paper was completed. Financial
support by DAAD, DFG, BMBF, and GSI is gratefully acknowledged.

\appendix
\section{Definition range for the YKP parameters}
\label{AS1}

The Yano-Koonin velocity $v$ in (\ref{3.2a}) is only defined 
if the discriminant
 \begin{equation}
   D = (A+B)^2 - 4 C^2 
 \label{a1}
 \end{equation}
is positive. Here we study the conditions under which this is the
case. To this end we write the expressions (\ref{3.3a})-(\ref{3.3c})
as
 \begin{eqnarray}
  A & = & A' - \frac1{\beta^2_\perp}
          \langle \tilde y^2 \rangle \, ,
 \label{a2} \\   
  B & = & B' - \frac{\beta^2_l}{\beta^2_\perp}
          \langle \tilde y^2 \rangle \, ,
 \label{a3} \\   
  C & = &  C' - \frac{\beta_l}{\beta^2_\perp}
          \langle \tilde y^2 \rangle \, .
 \label{a4}      
 \end{eqnarray}
We then have
 \begin{eqnarray}
  A' & = & \left \langle \left (\tilde t - \frac1{\beta_\perp} 
     \tilde x \right )^2 
     \right \rangle \ge 0
 \label{a5} \\
  B' & = & \left \langle \left (\tilde z - 
  \frac{\beta_l}{\beta_\perp} \tilde x \right )^2 
  \right \rangle \ge 0
 \label{a6} \\
  C' & = & \left \langle \left (\tilde t - 
       \frac1{\beta_\perp} \tilde x \right )
       \left (\tilde z - 
       \frac{\beta_l}{\beta_\perp} \tilde x \right ) \right \rangle
 \label{a7}
 \end{eqnarray}
and thus
 \begin{equation}
  |A'+B'| = A' + B' \, .
 \label{a8}
 \end{equation}
We will prove that
 \begin{equation}
  (A' + B')^2 - 4{C'}^2 \ge 0
 \label{a9}
 \end{equation}
or
 \begin{equation}
  |A'+B'| - 2 |C'| \ge 0 \, .
 \label{a10}
 \end{equation}
Due to (\ref{a8}) the inequality (\ref{a10}) can be written as
 \begin{equation}
 \label{a11}
  A' + B' \mp 2C' \ge 0
 \end{equation}
where the upper sign stands for $C'>0$ and the lower for $C'<0$.
Inserting expressions (\ref{a5})-(\ref{a7}) we obtain
 \begin{equation}
 \label{a12}
  A' + B' \mp 2 C' = \left \langle { \left \{ 
     \left (\tilde t - \frac1{\beta_\perp} \tilde x \right ) \mp
     \left (\tilde z - \frac{\beta_l}{\beta_\perp} \tilde x \right ) \right \} 
     }^2 \right \rangle
     \ge 0 \, .
 \end{equation}
This proves (\ref{a9}).
Since $A'$, $B'$, and $C'$ are the shorthands belonging to the modified
YKP parametrization, we have also proven that this parametrization is defined 
everywhere (except, of course, the point $K_\perp = \beta_\perp = 0$
\cite{WHTW96,HTWW96a}).

From expressions (\ref{a2})-(\ref{a4}) we see that
 \begin{eqnarray}
 \label{a13}
  A + B & = & A' + B' - \frac{1 + \beta_l^2}{\beta_\perp^2} 
    \langle \tilde y^2 \rangle \\ 
 \label{a14}
  2 C & = & 2 C' - \frac{2 \beta_l}{\beta_\perp^2} \langle \tilde y^2 \rangle 
  \, .
 \end{eqnarray}
It is therefore possible to obtain negative values for $D$:
 \begin{equation}
 \label{a15}
  D = (A+B)^2 - 4C^2 < 0 \, .
 \end{equation}
For example, if $\beta_l = 0$ and $\langle \tilde y^2 \rangle$ is fixed
at a value bigger than $\langle \tilde x^2 \rangle$, one can find
values for $\beta_\perp$ such that
 \begin{equation}
 \label{a16}
  |A+B| = \left | A' + B' - \frac{1}{\beta_\perp^2}
    \langle \tilde y^2 \rangle \right | < 2 |C'| = 2 |C|\, .
 \end{equation}
We must therefore conclude that there are kinematical regions where
the YKP para\-me\-tri\-za\-tion is not defined.

\begin{flushleft}

\end{flushleft}

\end{document}